\documentclass[12pt]{iopart}
\usepackage{amssymb}
\usepackage{amsfonts}
\usepackage{graphicx}
\usepackage{amsthm}

\renewcommand{\Re}{{\rm Re}}
\renewcommand{\Im}{{\rm Im}}
\newcommand{\C}{{\mathbb C}}
\newcommand{\R}{{\mathbb R}}
\newcommand{\HH}{{\cal H}}
\newcommand{\EE}{{\cal E}}
\newcommand{\TT}{{\cal T}}
\newcommand{\LLL}{{\cal L}}
\newcommand{\ket}[1]{|{#1}\rangle}
\newcommand{\kb}[1]{|{#1}\rangle \langle{#1}|}

\newcommand{\bra}[1]{\langle{#1}|}
\newcommand{\bkt}[2]{\langle{#1}|{#2}\rangle}
\newcommand{\kbt}[2]{|{#1}\rangle \langle{#2}|}
\newcommand{\Ref}[1]{(\ref{#1})}
\newcommand{\wv}[2]{\langle{#1}\rangle_{#2}}

\makeatletter

\newtheorem{thm}{Theorem}

\begin{document}
\title[Weak Values with Decoherence]
	{Weak Values with Decoherence}
 \author{Yutaka Shikano$^{1,2,\ast}$ and Akio Hosoya$^{1,\dagger}$}
 \address{$^1$ Department of Physics, Tokyo Institute of Technology, Meguro, Tokyo 152-8551, Japan}
 \address{$^2$ Department of Mechanical Engineering, Massachusetts Institute of Technology, Cambridge, MA 02139, USA}
 \ead{$^\ast$\mailto{shikano@mit.edu}, $^\dagger$\mailto{ahosoya@th.phys.titech.ac.jp}}
 \date{\today}
\begin{abstract}
The weak value of an observable is experimentally accessible 
by weak measurements as theoretically analyzed by Aharonov {\it et al.} and recently
experimentally demonstrated. We introduce a W operator associated with the weak values and give a general framework
of quantum operations to the W operator in parallel with the Kraus representation
of the completely positive map for the density operator. The decoherence effect is also investigated in 
terms of the weak measurement by a shift of a probe wave function of continuous variable. As an application,
we demonstrate how the geometric phase is affected by the bit flip noise.
\end{abstract}
\submitto{\JPA}
\maketitle
\section{Introduction}
The concept of measurement has been a longstanding problem in quantum mechanics. It was discussed by von Neumann~\cite{NEUMANN} in the mathematical foundations of quantum mechanics and further developed by Kraus~\cite{KRAUS}, Davies and Lewis~\cite{DL}, Ozawa~\cite{OZAWA84}, and other people (e. g., see~\cite{BML,BK}). According to the extended Born rule, the positive operator valued measure (POVM) $M_i$ gives the probability distribution $P_i= \Tr[\rho M_i]$ in quantum measurement with $\rho$ being the density operator corresponding to a prepared state. Measurement changes the quantum state as a positive map from a density operator $\rho(> 0)$ to another density operator $\EE(\rho)(> 0)$ because of the probabilistic interpretation. The map $\EE$  is called a completely positive map (CP map) if the map remains positive when the map is trivially extended to any larger Hilbert space. That is, $(\EE \otimes I)(\rho_{ex})>0$ for any state $\rho_{ex}$ of an arbitrarily extended system when the map $\EE$ is completely positive. Physically, this is a very reasonable requirement because there should always exist the outside of an experimental set up which is inactive during the experiment procedures~\cite{OZAWA84}. 
 
It is known that the seemingly humble requirement of the complete positivity of the quantum operation $\EE$ together with the trace preservation and the positive convexity for the density operator $\rho$ implies an explicit representation of the physical operation in the Kraus form: $\EE(\rho)=\sum_i E_{i} \rho E_{i} ^{\dagger}$, where $E_i$ is called the Kraus operator and $M_i := E^{\dagger}_i E_i $ is the POVM, with the property of the decomposition of unity, $\sum_{i} E_{i}^{\dagger}E_{i}=\sum_{i}M_i=1$. The Kraus representation of physical operations is a  powerful tool in quantum information theory~\cite{NC}.

However, the probability distribution is not the only thing that is experimentally accessible in quantum mechanics. In quantum mechanics, the phase is also an essential ingredient and in particular the geometric phase is a notable example of an experimentally accessible quantity~\cite{SW}. The general experimentally accessible quantity which contains complete information of the probability and the phase seems to be the {\it weak value} advocated by Aharonov and his collaborators ~\cite{AAV,AR}. They proposed a model of weakly coupled system and probe (See Sec. \ref{WMR}.) to obtain information to a physical quantity as a ``weak value" only slightly disturbing the state. Here, we briefly review the formal aspects of the weak value.

{\it Weak Value}: 
For an observable $A$, the weak value $\wv{A}{w}$ is defined as 
\begin{equation}
\wv{A}{w} = \frac{{\bra{f}}U(t_f,t)AU(t,t_i)\ket{i}}{\bra{f}U(t_f,t_i)\ket{i}}\in \C,
\end{equation} 
where $\ket{i}$ and $\bra{f}$ are normalized pre-selected ket and post-selected bra state vectors, respectively. Here, $U(t_2,t_1)$ is an evolution operator from the time $t_1$ to $t_2$. The weak value
$\wv{A}{w}$ actually depends on the pre- and post-selected states $\ket{i}$ and $\bra{f}$ but we omit them for notational simplicity in the case that we fix them. Otherwise, we write them explicitly as $_{f}\wv{A}{i}^{w}$ instead for $\wv{A}{w}$. The denominator is assumed to be non-vanishing. Note also that the weak value $\wv{A}{w}$ is independent of the phases of the pre- and post-selected states so that it is defined in the ray space. 

The physical intuition may be enhanced by looking at the identity for the expectation value of an observable $A$,

\begin{eqnarray}
\bra{{i}}U^{\dagger}(t,t_i)AU(t,t_i)\ket{i}
&=\sum_{f}|\bra{f}U(t_f,t_i)\ket{i}|^{2} \frac{{\bra{f}}U(t_f,t)AU(t,t_i)\ket{i}}{\bra{f}U(t_f,t_i)\ket{i}} \nonumber \\
&=\sum_{f}p_{f} \cdot _{f}\wv{A}{i}^{w},
\end{eqnarray}
where $p_{f}=|\bra{f}U(t_f,t_i)\ket{i}|^{2}$ is the probability to obtain the final state $\bra{f}$ given the initial state $\ket{i}$~\cite{AB}. Comparing with the standard probability theory, one may interpret the weak value as a complex random variable with the probability measure $p_{f}$~\footnote{	In the standard probability theory~\cite{ITO}, the expectation value of an observable $A$ is given as a probabilistic average, 
	$ \wv{A}{} = \int dp h_A (p)$, with $h_A(p)$ and $dp$ being the random variable associated with $A$ and probability 
	measure, which is independent of $A$. The standard expression of the expectation value,  
	$\wv{A}{} = \sum_n a_n |\bkt{a_n}{i}|^2$, is given by the Born rule. 
	However, this does not fit the standard probability theory because the probability measure depends on $A$.
}. The statistical average of the weak value coincides with the 
expectation value in quantum mechanics. Further, if an operator $A$ is a projection operator $A=\kb{a}$, the above identity 
becomes an analog of the Bayesian formula,
\begin{equation}
|\bra{a}U(t,t_i)\ket{i}|^2=\sum_{f}p_{f} \cdot _{f}\wv{\ket{a}\bra{a}}{i}^{w}.
\label{bay}
\end{equation} 
The left hand side is the probability to obtain the state $\ket{a}$ given the initial state $\ket{i}$.
From this, one may get some intuition by interpreting the weak value $_{f}\wv{\ket{a}\bra{a}}{i}^{w}$ as the (complex!) conditional probability of obtaining the result $\ket{a}$ under an initial condition $\ket{i}$ and a final condition $\ket{f}$ in the process $\ket{i} \rightarrow \ket{a} \rightarrow \ket{f}$~\footnote{The interpretation of the weak value as a complex probability is suggested in the literature \cite{MJP}.}.
We believe that the concept of a quantum trajectory can be formulated in the framework of weak values~\cite{WISEMAN}.
This interpretation of the weak values gives many possible examples of strange phenomena like a negative kinetic energy~\cite{APRV}, a spin $100 \hbar$ for an electron~\cite{AAV} and a superluminal propagation of light~\cite{RY}. Of course, we should not take the strange weak values too literally but the remarkable consistency of the framework of the weak values due to Eq. (\ref{bay}) and a consequence of the completeness relation,
\begin{equation}
\sum_{a}\wv{\ket{a}\bra{a}}{w}=1,
\label{comp}
\end{equation}
may give a useful concept to further push theoretical consideration by intuition. The framework of weak values has been theoretically applied to quantum stochastic process~\cite{WANG}, the tunneling traverse time~\cite{STEINBERG}, non-locality and consistent history~\cite{VAIDMAN}, semi classical weak values~\cite{TANAKA}, counterfactual reasonings~\cite{MELMER,MJP}, and quantum communications~\cite{BACGS}. However, the most important fact is that the weak value is experimentally accessible by the weak measurement (e. g., see \cite{RSH,RLS,RESCH,PCS,HK,Pryde,YYKI, Dixon}) so that the intuitive argument based on the weak values can be either verified or falsified by experiments.
Historically, the terminology ``weak value" comes from the weak measurement, where the coupling between the target system and the probe is weak, explained in Sec. \ref{WMR}. Below, we introduce a formal concept of a W operator as a tool to formally describe the weak value. 

{\it W operator}: 
We define a {\it W operator} $W(t)$ as
\begin{equation}
W(t) := U(t,t_i)\ket{i}{\bra{f}}U(t_f,t).
\label{definition}
\end{equation} 
To facilitate the formal development of the weak value, we introduce the ket state $\ket{\psi(t)}$ and the bra state $\bra{\phi(t)}$ as
\begin{equation}
	\ket{\psi(t)}= U(t,t_i)\ket{i} , \ \bra{\phi(t)} =\bra{f} U(t_f,t),
\end{equation}
so that the expression for the W operator simplifies to
\begin{equation}
W(t) = \kbt{\psi(t)}{\phi(t)}.
\label{tanweak}
\end{equation} 
By construction, the two states $\ket{\psi(t)}$ and $\bra{\phi(t)}$ satisfy the Schr{\"o}dinger equations with the same Hamiltonian with 
the initial and final conditions $\ket{\psi(t_i)}=\ket{i}$ and $\bra{\phi(t_f)}=\bra{f}$. In a sense, $\ket{\psi(t)}$ evolves forward in time while $\bra{\phi(t)}$ evolves backward in time. The time reverse of the W operator (\ref{tanweak}) is $W^{\dagger} = \kbt{\phi (t)}{\psi (t)}$. Thus, we can say the W operator is based on the two-state vector formalism~\cite{AV90}. Historically speaking, the two-state vector formalism was originally motivated by the time symmetric description of quantum measurement~\cite{ABL} and has been related to the weak values and weak measurement~\cite{AV91} also developed by Aharonov {\it et al.}~\cite{AV08,AT}. Even an apparently similar quantity to the W operator (\ref{tanweak}) was introduced by Reznik and Aharonov~\cite{RA} in the name of ``two-state" with the conceptually different meaning. The W operator gives the weak value of the observable $A$ as
\begin{equation}
	\wv{A}{w} = \frac{\Tr (WA)}{\Tr W}, 
\end{equation}
in parallel with the expectation value of the observable $A$ by $\Tr (\rho A) / \Tr \rho$ from Born's probabilistic interpretation. Furthermore, the W operator (\ref{definition}) can be regarded as a special case of a standard purification of the density operator~\cite{UHLMANN86}. In our opinion, the W operator should be considered on the same footing of the density operator. For a closed system, both satisfy the Schr{\"o}dinger equation. In a sense, the W operator $W$ is the square root of the density operator since 
\begin{equation}
	W(t)W^{\dagger}(t) = \ket{\psi(t)}\bra{\psi(t)} = U(t,t_i) \kb{i} U^{\dagger}(t,t_i),
\end{equation} 
which describes a state evolving forward in time for a given initial state $\ket{\psi(t_i)}\bra{\psi(t_i)}=\ket{i}\bra{i}$, while
\begin{eqnarray}
	W^{\dagger}(t)W(t) = \ket{\phi(t)}\bra{\phi(t)} = U(t_f,t) \kb{f} U^{\dagger}(t_f,t),
\end{eqnarray} 
which describes a state evolving backward in time for a given final state $\ket{\phi(t_f)}\bra{\phi(t_f)}=\ket{f}\bra{f}$. The W operator describes the entire history of the state from the past ($t_i$) to the future ($t_f$) and measurement performed at the time $t$ as we shall see in Sec. \ref{WMR}. This description is conceptually different from the conventional one by the time evolution of the density operator.

Our aim is to find the most general map for the W operator $W$. The result turns out to be of the form $\EE(W)=\sum_m E_m W F^{\dagger}_m $~\footnote{Of course, if we introduce a probe Hilbert space and then consider the probability distribution of the measurement 
	outcome, we can extract information of the phase by an interference pattern (e. g., see \cite{TONOMURA89,ARNDT}). 
	The virtue of the CP map is that it is defined solely by operations of the target Hilbert space. We would like to find 
	out a representation analogous to the Kraus representation for the phase related object in a way defined only in the target system. 
	A special case of weak values in the environment has recently been discussed on the post-selected 
	decay rate in \cite{Davies}.
}. The W operator is a useful tool to compactly describe the effect of decoherence to the weak values just as the density operator is to the expectation value in the standard theory of decoherence~\cite{GZ}. 
Furthermore, the amount of the effect due to the environment in the weak measurement is exactly given by the weak value defined by the quantum operation of the W operator $\EE(W)$.

This paper is organized as follows. In Sec. \ref{QOFQS}, we recapitulate the CP map as quantum operations for density operators. We define a W operator motivated by the two-state vector formalism and construct quantum operations for the W operator instead of the density operator. Because of the complete positivity of the quantum operation, we can construct the state change defined by operations solely on the target Hilbert space similarly to the Kraus representation. In Sec. \ref{WM}, we review the process to obtain weak values by the weak measurement~\cite{AAV,JOZSA}. We analyze the target system with the noise during the weak measurement by the shifts of the probe observables. In Sec. \ref{GF}, we show an application of the quantum operation to the W operator. Since the geometric phase can be formally given by weak values~\cite{Sjoqvist}, we operationally define the geometric phase for mixed states on the basis of the quantum operation of the W operator. Section \ref{SD} is devoted to the summary. 
\section{Quantum Operations} \label{QOFQS}
\subsection{Quantum Operations for Density Operators --- Review}
Let us recapitulate the general theory of quantum operations of a finite dimensional quantum system~\cite[Ch. 8]{NC}. All physically realizable quantum operations can be generally described by a CP map~\cite{OZAWA84,OZAWA89}, since the isolated system of a target system and an auxiliary system always undergoes the unitary evolution according to the axiom of quantum mechanics~\cite{NEUMANN}. One of the important properties of the CP map is that all physically realizable quantum operations can be described only by operators defined in the target system. 

Let $\EE$ be a positive map from $\LLL(\HH_s)$, a set of linear operations on the Hilbert space $\HH_s$, to $\LLL(\HH_s)$. If $\EE$ is completely positive, its trivial extension $\sigma$ from $\LLL(\HH_s)$ to $\LLL(\HH_s \otimes \HH_e)$ is also positive such that 
\begin{equation}
	\sigma (\ket{\alpha}) := (\EE \otimes I) (\kb{\alpha}) > 0,
	\label{sig}
\end{equation}
for an arbitrary state $\ket{\alpha} \in \HH_s \otimes \HH_e$. We assume without loss of generality ${\rm dim}\HH_s = {\rm dim}\HH_e < \infty$. Throughout this paper, we concentrate on the case that the target state is pure though the generalization to mixed states is straightforward. From the complete positivity, we obtain the following theorem for quantum state changes.
\begin{thm}
For any quantum state $\ket{\psi}_s \in \HH_s$, we expand 
\begin{equation}
	\ket{\psi}_s = \sum_m \psi_m \ket{m}_s,
\end{equation}
with a fixed complete orthonormal set $\{ \ket{m}_s \}$. Then, a quantum state change can be written as 
\begin{equation}
	\EE (\ket{\psi}_s \bra{\psi})= \, _e\bra{\tilde{\psi}} \sigma (\ket{\alpha}) \ket{\tilde{\psi}}_e,
	\label{schmidt}
\end{equation}
where
\begin{equation}
	\ket{\tilde{\psi}}_e := \sum_{m} \psi^{\ast}_{m} \ket{m}_{e}.
	\label{nishi}
\end{equation}
Here, $\ket{\alpha}$ is a maximally entangled state defined by 
\begin{equation}
	\ket{\alpha} := \sum_m \ket{m}_s \ket{m}_e,
	\label{sc}
\end{equation}
where $\{ \ket{m}_e \}$ is a complete orthonormal set corresponding to $\{ \ket{m}_s \}$. 
Equation (\ref{sig}) holds for the particular choice (\ref{sc}).
\label{1th}
\end{thm}
\proof{
We rewrite the right hand sides of Eq. \Ref{schmidt} as 
\begin{equation}
	\sigma (\ket{\alpha}) = (\EE \otimes I) \left( \sum_{m,n} \ket{m}_{s} \ket{m}_{e} \, _{s}\bra{n} _{e}\bra{n} \right) = \sum_{m,n} \ket{m}_{e} \bra{n} \EE (\ket{m}_s \bra{n}),
\end{equation}
to obtain
\begin{equation}
	_{e}\bra{m} \sigma (\ket{\alpha}) \ket{n}_{e} = \EE (\ket{m}_{s} \bra{n}).
\end{equation}
By linearity, we arrive at the desired equation (\ref{schmidt}). \qed}

From the complete positivity, $\sigma (\ket{\alpha}) > 0$, for the particular case (\ref{sc}), we can express $\sigma (\ket{\alpha})$ as 
\begin{equation}
	\sigma (\ket{\alpha}) = \sum_{m} s_{m} \kb{\hat{s}_{m}} = \sum_{m} \kb{s_{m}},
	\label{complete}
\end{equation}
where $s_m$'s are positive and $\{ \ket{\hat{s}_m} \}$ is a complete orthonormal set with $\ket{s_m}:=\sqrt{s_m} \ket{\hat{s}_m}$. 
We define the Kraus operator $E_m$~\cite{KRAUS} as 
\begin{equation}
	E_{m} \ket{\psi}_s := \, _{e}\bkt{\tilde{\psi}}{s_m},
	\label{krausreview}
\end{equation}
where $\ket{\tilde{\psi}}_e$ is defined in Eq. (\ref{nishi}).
Then, the quantum state change becomes the Kraus form, 
\begin{eqnarray}
	\sum_{m} E_m \ket{\psi}_s \bra{\psi} E^{\dagger}_{m} & = \sum_{m} \, _{e}\bkt{\tilde{\psi}}{s_m} \bkt{s_m}{\tilde{\psi}}_{e}
	= \, _{e}\bra{\tilde{\psi}} \sigma \ket{\tilde{\psi}}_{e} \nonumber \\
	& = \EE (\ket{\psi}_{s} \bra{\psi}).
\end{eqnarray}
We emphasize that the quantum state change is described solely in terms of the quantities of the target system.
\subsection{Quantum Operations for W operators}
Let us now define a W operator as 
\begin{equation}
	W(t) := \kbt{\psi (t)}{\phi (t)},
	\label{defwo}
\end{equation}
based on the two-state vector formalism by Aharonov and Vaidman~\cite{AV90} and define
\begin{equation}
\wv{A}{W} := \frac{\Tr (AW)}{\Tr (W)},
\end{equation}
for an observable $A$ corresponding to the weak value of the observable $A$~\cite{AAV} as the above~\footnote{While the original notation of the weak values is $\wv{A}{w}$ indicating the ``w"eak value of an observable 
	$A$, our notation is motivated by one of which the pre- and post-selected states are explicitly shown as 
	$_{f} \wv{A}{i}^{w}$.}. The weak value is an analog of a probability, and so is the W operator that of the density operators. We discuss a state change in terms of the W operator and define a map $X$ as 
\begin{equation}
	X (\ket{\alpha}, \ket{\beta}) := (\EE \otimes I) \left( \kbt{\alpha}{\beta} \right),
	\label{weakop}\\
\end{equation}
for an arbitrary $\ket{\alpha}, \ket{\beta} \in \HH_s \otimes \HH_e$. Then, we obtain the following theorem on the change of the W operator such as Theorem \ref{1th}.
\begin{thm}
For any W operator $W = \ket{\psi (t)}_{s} \bra{\phi (t)}$, we expand 
\begin{equation}
	\ket{\psi (t)}_{s} = \sum_{m} \psi_{m} \ket{\alpha_m}_{s}, \ \ket{\phi (t)}_{s} = \sum_{m} \phi_{m} \ket{\beta_m}_{s}, 
\end{equation}
with fixed complete orthonormal sets $\{\ket{\alpha_m}_{s} \}$ and $\{\ket{\beta_m}_{s} \}$.
Then, a change of the W operator can be written as  
\begin{equation}
	\EE \left( \ket{\psi (t)}_{s} \bra{\phi (t)} \right)= \, _{e}\bra{\tilde{\psi} (t)} X (\ket{\alpha}, \ket{\beta}) \ket{\tilde{\phi} (t)}_{e},
	\label{tvwo}
\end{equation}
where
\begin{equation}
	\ket{\tilde{\psi} (t)}_{e} = \sum_{k} \psi^{\ast}_{k} \ket{\alpha_k}_{e}, \ 
	\ket{\tilde{\phi} (t)}_{e} = \sum_{k} \phi^{\ast}_{k} \ket{\beta_k}_{e}, \label{tildetwo}
\end{equation}
and $\ket{\alpha}$ and $\ket{\beta}$ are maximally entangled states defined by 
\begin{equation}
	\ket{\alpha} := \sum_m \ket{\alpha_m}_s \ket{\alpha_m}_e, \ 
	\ket{\beta} := \sum_m \ket{\beta_m}_s \ket{\beta_m}_e.
\end{equation}
Here, $\{\ket{\alpha_m}_{e} \}$ and $\{\ket{\beta_m}_{e} \}$ are complete orthonormal sets corresponding to $\{\ket{\alpha_m}_{s} \}$ and $\{\ket{\beta_m}_{s} \}$, respectively.
\label{2th}
\end{thm}
The proof is completely parallel to that of Theorem \ref{1th}.

We take the polar decomposition of the map $X$ to obtain
\begin{equation}
	X = \sigma u,
\end{equation}
noting that 
\begin{equation}
	X X^{\dagger} = \sigma u u^{\dagger} \sigma = \sigma^2.
\end{equation}
The unitary operator $u$ is well-defined on $\HH_s \otimes \HH_e$ because $\sigma$ defined in Eq. (\ref{sig}) is positive. This is a crucial point to obtain the main result of this paper (\ref{wk}), which is the operator-sum representation for the quantum operation of the W operator.  
From Eq. (\ref{complete}), we can rewrite $X$ as 
\begin{eqnarray}
	X & = \sum_{m} \kb{s_m} u \nonumber \\
	& = \sum_{m} \kbt{s_m}{t_m},
\end{eqnarray}
where
\begin{equation}
	\bra{t_{m}} = \bra{s_m} u.
\end{equation}
Similarly to the Kraus operator (\ref{krausreview}), we define the two operators, $E_m$ 
and $F^{\dagger}_{m}$, as 
\begin{eqnarray}
	E_{m} \ket{\psi (t)}_s & := \, _{e}\bkt{\tilde{\psi} (t)}{s_m}, \\
	_{s}\bra{\phi (t)} F^{\dagger}_{m} & := \bkt{t_{m}}{\tilde{\phi} (t)}_{e},
\end{eqnarray}
where $\ket{\tilde{\psi} (t)}_e$ and $\ket{\tilde{\phi} (t)}_e$ are defined in Eq. (\ref{tildetwo}).
Therefore, we obtain the change of the W operator as 
\begin{eqnarray}
	\sum_{m} E_{m} \ket{\psi (t)}_s \bra{\phi (t)} F^{\dagger}_{m} & = \sum_{m} \, _{e}\bkt{\tilde{\psi}(t)}{s_m} \bkt{t_{m}}{\tilde{\phi}(t)}_{e} 
	= \, _{e}\bra{\tilde{\psi}(t)} X \ket{\tilde{\phi}(t)}_{e} \nonumber \\
	& = \EE \left( \ket{\psi (t)}_s \bra{\phi (t)} \right),
\end{eqnarray}
using Theorem \ref{2th} in the last line. By linearity, we conclude
\begin{equation}
\EE(W)=\sum_{m} E_{m}WF^{\dagger}_{m}, 
\label{wk}
\end{equation}
which is one of our main results. Note that, in general, $\EE (W) \EE(W^{\dagger}) \neq \EE(\rho)$ although $\rho = WW^{\dagger}$.

Summing up, we have introduced the W operator (\ref{defwo}) and obtained the general form of the quantum operation of the W operator (\ref{wk}) in an analogous way to the quantum operation of the density operator assuming the complete positivity of the physical operation.  
 
It is well established that the trace preservation, $\Tr(\EE(\rho))= \Tr \rho = 1$ for all $\rho$, implies that $\sum_{m}E^{\dagger}_{m}E_{m}=1$.
 The proof is simple \cite{NC} and goes through as 
 \begin{equation}
1 =\Tr(\EE(\rho)) = \Tr \left( \sum_{m}E_{m}\rho E^{\dagger}_{m} \right) = \Tr \left( \sum_{m}E^{\dagger}_{m}E_{m}\rho \right) \; ( \forall \rho ).
\end{equation}
This argument for the density operator $\rho= WW^{\dagger}$ applies also for $W^{\dagger} W$ to obtain $\sum_{m}F^{\dagger}_{m}F_{m}=1$ because this is the density operator in the time reversed world in the two-state vector formulation as reviewed in the introduction. Therefore, we can express the Kraus operators, 
\begin{equation}
E_{m} = \, _e\bra{e_m}U\ket{e_i}_e, \ F_{m}^{\dagger} = \, _e\bra{e_f}V\ket{e_m}_e, \label{kraus2}
\end{equation}
where 
\begin{equation}
	U = U (t, t_i), \ V = U(t_f,t), \label{twounitary}
\end{equation}
are the evolution operators, which act on $\HH_s \otimes \HH_e$. $\ket{e_i}$ and $\ket{e_f}$ are some basis vectors and  $\ket{e_m}$ is a complete set of basis vectors with $\sum_{m} \kb{e_m} = 1$. We can compute 
\begin{equation}
\sum_{m}F^{\dagger}_{m}E_{m} = \sum_{m} \, _e\bra{e_f}V\ket{e_m}_e \bra{e_m}U\ket{e_i}_e 
= \, _e\bra{e_f}VU\ket{e_i}_e.
\label{Sma}
\end{equation}
The above equality (\ref{Sma}) may be interpreted as a decomposition of the history in analogy to the decomposition of unity because
\begin{equation}
	\,_e\bra{e_f}VU\ket{e_i}_e = \, _e\bra{e_f}S\ket{e_i}_e = S_{fi} 
	\label{smatrix}
\end{equation}
is the S-matrix element.
The meaning of the basis $\ket{e_i}$ and $\ket{e_f}$ will be clear in the following section.

\begin{figure*}[t]
	\centering
	\includegraphics[width=15cm]{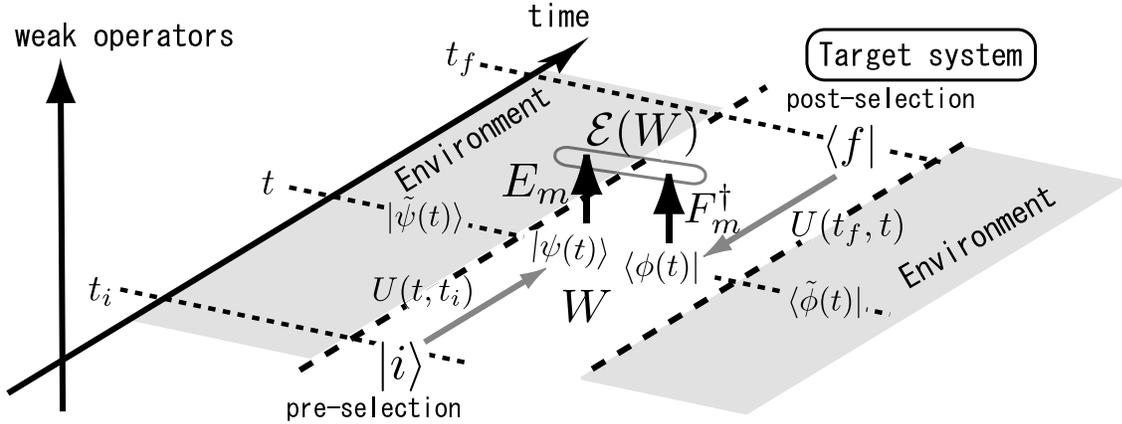}
	\caption{The quantum operations for the W operators. The W operator $W(t) = \ket{\psi (t)}_s \bra{\phi (t)}$ carries 
	the entire history from the pre-selected state to the post-selected state. The quantum 
	operations for the W operator is described by the two operators $E_m$ and $F^{\dagger}_m$. 
	These operators correspond to the Kraus operators for the density operators, $WW^{\dagger}$ and $W^{\dagger}W$, 
	related to the two-state vector formalism and affect the history.}
	\label{two}
\end{figure*}

As is well known~\cite{NC}, the physical operation for the density operator can also be described by introducing an environment which is tensored by the target system. We perform a unitary transformation for a combined state and then take a partial trace over the environmental states. We can also apply this method to the W operator. Namely,
 \begin{equation}
\EE(W)=\Tr_{env}[U(W\otimes e)V],
\end{equation}
where $e=\kbt{e_i}{e_f}$ is the environmental W operator before the physical process. It is straightforward to formally carry out the partial trace to reproduce the Kraus representation for the W operator $W$ as Eq. (\ref{wk}). Any interaction model of this type will give the same Kraus representation for the W operator. The procedure of quantum operations for the W operator is illustrated in Fig. \ref{two}. 

A representation of an ensemble of quantum states is usually described by a density operator since we only obtain the probability distribution of an observable in the conventional quantum measurement theory. The density operator does not contain the phase information of a quantum state. However, the W operator gives a weak value of an observable and retains the information of the phase of the quantum state. We will show a typical example of a geometric phase in Sec. \ref{GF}.
\section{Weak Measurement with Decoherence} \label{WM}
So far we have formally discussed the quantum operations of the W operators. In this section, we would like to study the effect of environment in the course of the weak measurement~\cite{AAV} and see how the shift of the probe position is affected by the environment. As we shall see, the shift is related to the quantum operation of the W operator $\EE(W)$ (\ref{wk}) which we have investigated in the previous section.
\subsection{Weak Measurement---Review} \label{WMR}
First, we recapitulate the idea of the weak measurement~\cite{AAV,JOZSA}. Consider a target system and a probe defined in the Hilbert space $\HH_s\otimes\HH_p$. The interaction of the target system and the probe is assumed to be weak and instantaneous,
\begin{equation}
H_{int} (t)=g\delta(t-t_0)(A\otimes P),
\end{equation}
where an observable $A$ is defined in $\HH_s$, while $P$ is the momentum operator of the probe. The time evolution operator becomes
 \begin{equation}
e^{-ig(A\otimes P)}.
\end{equation}
Suppose the probe state is initially $\xi(q) \in \R$ in the coordinate representation with the probe position $q$.
For the transition from the pre-selected state $\ket{i}$ to the post-selected state $\ket{f}$, the probe wave function becomes
 \begin{equation}
\bra{f}Ve^{-ig(A\otimes P)}U\ket{i} \xi(q),
\end{equation}
which is in the weak coupling case~\footnote{$\xi \left(q - g \frac{\bra{f}VAU\ket{i}}{\bra{f}VU\ket{i}} \right)$ stands for $\xi (q)|_{q \to 
	q - g \frac{\bra{f}VAU\ket{i}}{\bra{f}VU\ket{i}}}$.},
\begin{eqnarray}
&\bra{f}V[1-ig(A\otimes P)]U\ket{i} \xi(q) \nonumber \\ 
&=\bra{f}VU\ket{i} \xi(q)-g\bra{f}VAU\ket{i} \xi^{\prime}(q) \nonumber \\
&\approx \bra{f}VU\ket{i} \xi \left( q-g\frac{\bra{f}VAU\ket{i}}{\bra{f}VU\ket{i}} \right).
\end{eqnarray}
In the previous notation, the argument of the wave function is shifted by
\begin{equation}
g\frac{\bra{f}VAU\ket{i}}{\bra{f}VU\ket{i}}=g\wv{A}{w}
\end{equation}
so that the shift of the expectation value is the real part of the weak value, $g \cdot \Re [\wv{A}{w}]$.  The shift of the momentum distribution can be similarly calculated to give $2 g\cdot Var(p) \cdot \Im [\wv{A}{w}]$,
where $Var(p)$ is the variance of the probe momentum before the interaction. Putting together, we can measure the weak value $\wv{A}{w}$ by observing the shift of the expectation value of the probe both in the coordinate and momentum
representations. The shift of the probe position contains the future information up to the post-selected state.
\subsection{Weak Measurement and Environment} \label{known}
\begin{figure*}[t]
	\centering
	\includegraphics[width=8cm]{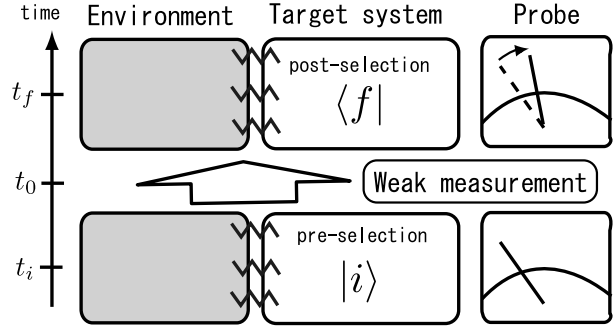}
	\caption{A weak measurement model with the environment. The environment affects the target system as a noise but does not 
	affect the probe. The weak measurement for the target system and the probe brings about the shift of the probe position 
	at $t_0$. The amount of the shift depends whether the environmental state is controllable (Sec. \ref{known}) or uncontrollable 
	(Sec. \ref{unknown}).}
	\label{environment}
\end{figure*}

Let us consider a target system coupled with an environment and a general weak measurement for the compound of the target system and the environment. We assume that there is no interaction between the probe and the environment. This situation is illustrated in Fig. \ref{environment}. The Hamiltonian for the target system and the environment is given by
\begin{equation}
H=H_0\otimes I_e+H_1,
\end{equation}
where $H_0$ acts on the target system $\HH_s$ and the identity operator $I_e$ is for the environment $\HH_e$, while $H_1$ acts on $\HH_s\otimes \HH_e $. 
The evolution operators $U=U(t,t_i)$ and $V=U(t_f,t)$ as defined in Eq. (\ref{twounitary}) can be expressed by
\begin{equation}
	U = U_{0} K(t_0,t_i), \ 
	V = K(t_f,t_0) V_{0},
\end{equation}
where $U_{0}$ and $V_{0}$ are the evolution operators forward in time and backward in time, respectively, by the target Hamiltonian $H_0$.
$K$'s are the evolution operators in the interaction picture,
\begin{equation}
	 K(t_0,t_i) = \TT e^{-i\int^{t_0}_{t_i} dt U_{0}^{\dagger} H_1 U_{0}}, \ 
	 K(t_f,t_0) = \overline{\TT} e^{-i\int^{t_f}_{t_0} dt V_{0} H_1 V_{0}^{\dagger}},
\end{equation}
where $\TT$ and $\overline{\TT}$ stand for the time-ordering and anti time-ordering products.

Let the initial and final environmental states be $\ket{e_i}$ and $\ket{e_f}$, respectively. The probe state now becomes
\begin{equation}
 \bra{f}\bra{e_f}VU\ket{e_i}\ket{i} \xi \left( q-g\frac{\bra{f}\bra{e_f}VAU\ket{e_i}\ket{i}}{\bra{f}\bra{e_f}VU\ket{e_i}\ket{i}} \right).
\end{equation}
Plugging the expressions for $U$ and $V$ into the above, we obtain the probe state as
\begin{equation}
N \xi \left( q-g\frac{\bra{f}\bra{e_f}K(t_f,t_0)V_{0}AU_{0}K(t_0,t_i)\ket{e_i}\ket{i}}{N} \right),
\end{equation}
where $N= \bra{f}\bra{e_f}K(t_f,t_0)V_{0}U_{0}K(t_0,t_i)\ket{e_i}\ket{i}$
is the normalization factor. We define the dual quantum operation as
\begin{eqnarray}
 \EE^{\ast}(A) & :=\bra{e_f}K(t_f,t_0)V_{0}AU_{0}K(t_0,t_i)\ket{e_i} \nonumber \\ 
 &=\sum_{m} V_{0} F^{\dagger}_{m}A E_m U_{0},
\end{eqnarray}
where 
\begin{equation}
	F^{\dagger}_{m} := V^{\dagger}_{0}\bra{e_f}K(t_f,t_0)\ket{e_m}V_0, \ 
	E_m := U_{0}\bra{e_m}K(t_0,t_i)\ket{e_i}U^{\dagger}_{0}
\end{equation}
are the Kraus operators introduced in the previous section (\ref{kraus2}). Here, we have inserted the completeness relation $\sum_{m}\ket{e_m}\bra{e_m}=1$ with $\ket{e_m}$ being not necessarily orthogonal. The basis $\ket{e_i}$ and $\ket{e_f}$ are the initial and final environmental states, respectively. This provides the meaning of $\ket{e_i}$ and $\ket{e_f}$ as alluded before. Thus, we obtain the wave function of
the probe as
\begin{eqnarray}
\xi \left( q-g\frac{\bra{f}\EE^{*}(A)\ket{i}}{N} \right) & = \xi \left( q - g \frac{\sum_{m} \bra{f}V_0 F^{\dagger}_{m} A E_{m} U_0 \ket{i}}{\sum_{m} \bra{f} V_0 F^{\dagger}_m E_m U_0 \ket{i}} \right) \nonumber \\
& = \xi \left( q - g \frac{\Tr \left[ A \sum_m E_m U_0 \kbt{i}{f} V_0 F^{\dagger}_m \right]}{\Tr \left[ \sum_m E_m U_0 \kbt{i}{f} V_0 F^{\dagger}_m \right]} \right) \nonumber \\
& = \xi \left( q-g\frac{\Tr[\EE(W)A]}{\Tr [\EE (W)]}\right) = \xi (q-g\wv{A}{\EE(W)}),
\end{eqnarray}
with $N= \bra{f}\EE^{*}(I)\ket{i}$ up to the overall normalization factor. This is the main result of this subsection. The shift of the expectation value of the position operator on the probe is
\begin{equation}
\delta q=g \cdot \Re[\wv{A}{\EE(W)}].
\end{equation} 
From an analogous discussion, we obtain the shift of the expectation value of the momentum operator on the probe as 
\begin{equation}
\delta p = 2 g \cdot Var(p) \cdot \Im[\wv{A}{\EE(W)}].
\end{equation}
Thus, we have shown that the probe shift in the weak measurement is exactly given by the weak value defined by the quantum operation of the W operator due to the environment, which is one of our main results.
\subsection{Weak Measurement---Decoherence} \label{unknown} 
In many cases, the initial and final states, $e_i$ and $e_f$, of the environment, on which the quantum operation $\EE$ depends, are not controllable so that they have to be statistically treated. Let the statistical weight be $w(e_f,e_i)$ and consider the average,
 \begin{equation}
Ave(h):=\sum_{e_f,e_i}w(e_f,e_i) h(e_f,e_i), 
\end{equation} 
for a function $h$ of the random variables $e_f$ and $e_i$. We proceed 
\begin{eqnarray}
& Ave \left( N \xi \left( q-g\frac{\Tr[\EE(W)A]}{N} \right) \right) \nonumber \\
& \approx  Ave \left( N \left\{ \xi \left( q \right) - g\frac{\Tr[\EE(W)A]}{N} \xi^{\prime} \left( q \right) \right\} \right) \nonumber \\
& = Ave(N) \left[ \xi(q) - g \frac{Ave [\Tr \EE (W) A]}{Ave (N)} \xi^{\prime} (q) \right] \nonumber \\
& \approx Ave( \Tr \EE (W)) \xi \left( q-g\frac{Ave(\Tr[\EE(W)A])}{Ave(\Tr[\EE(W)])} \right),
\end{eqnarray}
noting that 
\begin{eqnarray}
	N & = \bra{f} \EE^{\ast} (I) \ket{i} = \sum_m \bra{f} V_0 F^{\dagger}_m E_m U_0 \ket{i} \nonumber \\
	& = \sum_m \Tr E_m U_0 \kbt{i}{f} V_0 F^{\dagger}_m = \Tr \EE (W).
\end{eqnarray}
We see that the shift of the expectation value of the probe position is on average,
\begin{equation}
\delta q= \Re \left[ g \frac{Ave(\Tr[\EE(W)A])}{Ave(\Tr[\EE(W)])} \right],
\end{equation}
in the weak coupling case.
To obtain a significant shift, one needs some prior knowledge of
the environment. For the case of a detector as the environment, $e_i$ and $e_f$ are specified by the measurement outcome and are definite if the environment is at zero temperature, for example. 
\section{Geometric Phase} \label{GF}
\begin{figure*}[t]
	\centering
	\includegraphics[width=10cm]{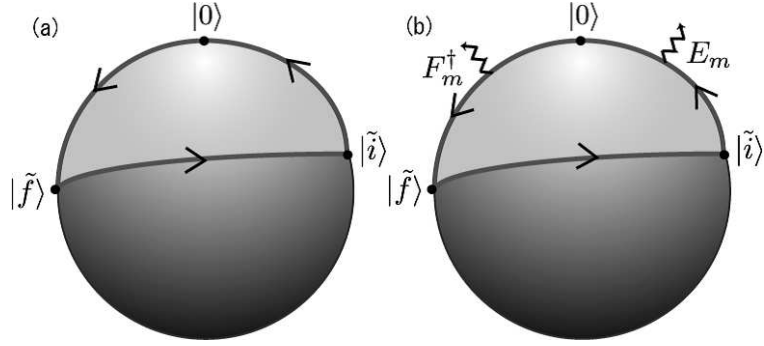}
	\caption{(a) Geometric phase in the case of the pure state for a qubit system. When the path 
	$\ket{\tilde{i}}\rightarrow \ket{P} \rightarrow \ket{\tilde{f}}$, the half of the solid 
	angle on the Bloch sphere, which is the gray region, corresponds 
	to the geometric phase. (b) Geometric phase in the presence of the environment with the same path. 
	The quantum operations are represented by $E_m$ and $F^{\dagger}_{m}$.}
	\label{phase}
\end{figure*}
We present a simple application of our framework of the physical operation of the W operators to the geometric phase since measuring the geometric phase is equivalent to measuring the weak value in the qubit system~\cite{Tamate}. There have been many works on the ``geometric phase for mixed states" but the very definition seems under controversy~\cite{Sjoqvist3, Ericsson, Carollo, Singh, Du, Ericsson2, Sjoqvist2}. We would like to start with the geometric phase $\gamma$ in a pure state~\cite{Sjoqvist} which is well-defined and can be expressed in terms of the weak value, 
\begin{equation}
	\gamma= \arg \left[ \frac{\Tr (WP)}{\Tr (W)} \right],
	\label{geopure}
\end{equation}
where $P = \kb{P}$ is a projector to a pure state and $W = \kbt{\tilde{i}}{\tilde{f}}$. Here, we simplify the notations 
$\ket{\tilde{i}} := U(t,t_i)\ket{i}$ and $\bra{\tilde{f}} := \bra{f}U(t_f,t)$ only in this section. 
The geometric phase $\gamma$ corresponds to the quantum path 
$\ket{\tilde{i}}\rightarrow \ket{P} \rightarrow \ket{\tilde{f}}$.
By the physical operation $\EE$, the W operator and the density operator are mapped to $\EE(W)$ and
$\EE(\rho)$, respectively. The new state $\EE(\rho)$ is in general a mixed state. The new geometric phase $\gamma_g$
 is correspondingly given by 
\begin{equation}
	\gamma_g = \arg \left[ \frac{\Tr(\EE(W)P)}{\Tr \EE(W)} \right],
	\label{geomix}
\end{equation}
which might be called the geometric phase of the mixed state
 $\EE(\rho)$ by a slight abuse of words. This operational definition fits well to experimental situation. That is,
 an experimentalist starts with a pure state $\ket{\tilde{i}}$ and then makes a trip 
$\ket{\tilde{i}}\rightarrow \ket{P} \rightarrow \ket{\tilde{f}}$ by manipulating
the external field. If there were no decoherence during that process, one would get the geometric phase defined above (\ref{geopure}). 
Otherwise, one would instead get the value (\ref{geomix}) for the geometric phase, while one can presume that the state 
is $\EE(\rho)=\EE(WW^{\dagger})$. Furthermore, we would like to point out that this definition (\ref{geomix}) coincides 
with the formal definition in the Uhlmann approach~\cite{UHLMANN86, Sjoqvist2}. In the generalized Kraus representation, the geometric 
phase for that path, $\ket{\tilde{i}}\rightarrow \ket{P} \rightarrow \ket{\tilde{f}}$ can be written as
\begin{equation}
	\gamma_g = \arg \left[\frac{\sum_{m} \bra{\tilde{f}}F^{\dagger}_{m}\ket{P} \bra{P}E_{m} \ket{\tilde{i}}}{\sum_m \bra{\tilde{f}} F^{\dagger}_m E_m \ket{\tilde{i}}} \right].
	\label{gpmix}
\end{equation}
This is illustrated in Fig. \ref{phase}. The expression (\ref{gpmix}) is operationally defined from the view of the decoherence and 
may be regarded as an improvement of Eq. (10) in the paper~\cite{Sjoqvist2}.

Let us see the decoherence effect on the geometric phase in a simple one qubit system under a bit flip noise \cite{NC}. The Kraus operators are chosen by
\begin{equation}
	E_{0} = F_{0} =  \sqrt{p} I, \ 
	E_{1} = F_{1} = \sqrt{1-p}\sigma_{x}, 
\end{equation}
with $0 \leq p \leq 1$ note that
\begin{equation}
	\sum_m F^{\dagger}_m E_m = I
\end{equation}
is the unitary operator and $\sum_m E^{\dagger}_m E_m = \sum_m F^{\dagger}_m F_m = I$.
We are going to consider a path, $\ket{\tilde{i}}\rightarrow \ket{0}\rightarrow \ket{\tilde{f}}$, where
\begin{equation}
	\ket{\tilde{i}} = \frac{1}{\sqrt{2}}( \ket{0}- \ket{1}), \ 
	\ket{\tilde{f}} =\frac{1}{\sqrt{2}}( \ket{0}- e^{-i \phi} \ket{1}),
\end{equation}
where $\ket{\tilde{f}}$ is a rotated state from $\ket{\tilde{i}}$ by the angle $\phi$.
A straightforward calculation shows 
\begin{equation}
	\wv{P_0}{\EE (W)} = \frac{p + (1-p) e^{i \phi}}{1 + e^{i \phi}}
\end{equation}
for the projection, $P_0 = \kb{0}$. Then, the geometric phase $\gamma_g$ (\ref{gpmix}) is 
\begin{equation}
	\gamma_g = \arg \wv{P_0}{\EE (W)} = \arctan \left[ (1-2p) \tan \frac{\phi}{2} \right] .
\end{equation}
In the no noise case, $p=1$, we recover the geometric phase $- \phi/2$, which is the half of the solid angle formed by the three vectors $\ket{\tilde{i}},\ket{0}$, and $\ket{\tilde{f}}$ in the Bloch sphere. In the case of $p=1/2$, the geometric phase vanishes, as we expect because the state is completely mixed. It seems the decohered geometric phase has no particular geometrical meaning while Uhlmann gave an expression for the geometric phase in terms of operators similar to the W operator~\cite{UHLMANN91,UHLMANN92}. It is curious to point out that the geometric phase during the measuring process is given by a time evolution of weak values~\cite{PARKS}.
 
We would like to stress that our definition of the geometric phase under the environmental noise is operationally defined in the sense that the geometric phase is initially defined in a pure state but undergoes a decoherence process while the state becomes a mixed state. Hence, the geometric phase of the mixed state (\ref{geomix}) can be experimentally verifiable.
\section{Summary} \label{SD}
We have introduced the W operator $W$ (\ref{definition}) to formally describe the weak value advocated by Aharonov {\it et al.} 
which is the more general quantity containing the phase information than the density operator $\rho$. The general framework 
is given to describe effects of quantum operation $\EE(W)$ to the W operator $W$ in parallel with the Kraus representation of the completely positive map for the density operator $\rho$. The result is the change of the history, 
\begin{equation}
	\EE(W) = \sum_m E_m W F^{\dagger}_m,
\end{equation}
with $\sum_m E^{\dagger}_m E_m = \sum_m F^{\dagger}_m F_m = 1$ and $\sum_m F^{\dagger}_m E_m = S_{fi}$ is the S-matrix element corresponding to the history (See Fig. \ref{two}). We have shown the effect of the environment during the weak measurement as the shift of the expectation value of the probe observables in both cases of the controllable and uncontrollable environmental states. As an application, it is exhibited how the geometric phase is affected by decoherence, which is experimentally testable. 

Extending our proposed definition of the W operators, we may consider a superposition of W operators, 
\begin{equation}
W := \sum_{i,f} \alpha_{if} U(t,t_i)\ket{i}\bra{f}U(t_f,t),
\label{mix}
\end{equation}
in analogy to the mixed state which is a convex linear combination of pure states. Actually, $\EE(W)$ (\ref{wk}) has the form (\ref{mix}). Although this indicates a time-like correlations, the physical implication is not yet clear. This operator may be related to the concept of the multi-time states~\cite{APTV}. In fact, it is shown how the weak value corresponding to the W operator (\ref{mix}) can be constructed via a protocol by introducing auxiliary states which are space-likely entangled with the target states. 

\section*{Acknowledgments}
We would like to acknowledge useful comments with Dr. Alberto Carlini, Professor Masanao Ozawa, and Professor Holger Hofmann and one of the authors (YS) would like to also acknowledge useful discussions with Professor Masao Kitano. We would like to thank the support from Global Center of Excellence Program ``Nanoscience and Quantum Physics" at Tokyo Institute of Technology. YS acknowledges JSPS Research Fellowships for Young Scientists (Grant No. 21008624).
\section*{References}

\end{document}